\documentclass[aps,prc,twocolumn,floatfix,showpacs,a4paper,
nofootinbib,amsmath,amssymb]{revtex4-1}
\usepackage{graphicx}
\usepackage{dcolumn}
\usepackage{bm}

\newcommand{\be}{\begin{equation}}
\newcommand{\ee}{\end{equation}}
\newcommand{\ba}{\begin{eqnarray}}
\newcommand{\ea}{\end{eqnarray}}
\newcommand{\bd}{\begin{displaymath}}
\newcommand{\ed}{\end{displaymath}}
\renewcommand{\vec}[1]{\mbox{\boldmath$#1$}}

\begin{document}
\title{Flow Vorticity in Peripheral High Energy Heavy Ion Collisions}

\author{L.P.~Csernai$^1$, V.K.~Magas$^2$, and D.J. Wang$^1$}

\affiliation{
$^1$ Institute of Physics and Technology, University of Bergen,
Allegaten 55, 5007 Bergen, Norway \\
$^2$ Departament d'Estructura i Constituents de la Mat\`eria,
Universitat de Barcelona, 08028 Barcelona, Spain
}

\begin{abstract}
The vorticity development is studied in the reaction plane of
peripheral relativistic heavy ion reactions where the initial
state has substantial angular momentum. The earlier predicted
rotation effect and Kelvin Helmholtz Instability, lead to
significant initial vorticity and circulation.
In low viscosity QGP this vorticity remains still significant
at the time of freeze out of the system, even if damping due
to the explosive expansion and the dissipation decreases the
vorticity and circulation. In the reaction plane the
vorticity arises from the initial angular momentum, and it
is stronger than in the transverse plane where vorticity is
caused by random fluctuations only.
\end{abstract}

\date{\today}

\pacs{89.65.-s, 89.75.Da, 05.45.Tp.}

\maketitle

\section{Introduction}
The strongly interacting quark gluon plasma (QGP) \cite{Gyulassy} has raised
many interesting questions about the physics of ultra dense hot matter
produced in relativistic heavy ion collisions. Many transport and
hydrodynamic models \cite{models} are popular in analyzing the evolvement
of the viscous QGP and its properties.
In peripheral heavy ion reactions due to the initial angular momentum the
initial state of the fluid dynamical stage of the collision dynamics has
a shear flow characteristics, and this leads to rotation \cite{hydro1}
and even Kelvin Helmholtz Instability (KHI) \cite{hydro2} in the reaction
plane, for low viscosity Quark-gluon plasma.  This possibility was indicated
by high resolution Computational Fluid Dynamics (CFD) calculations using
the PIC method. We study the development of these processes in 3+1
dimensional (3+1D) configuration to describe the energy and momentum
balance realistically.
\begin{figure}[ht]  
\begin{center}
\resizebox{0.95\columnwidth}{!}
{\includegraphics{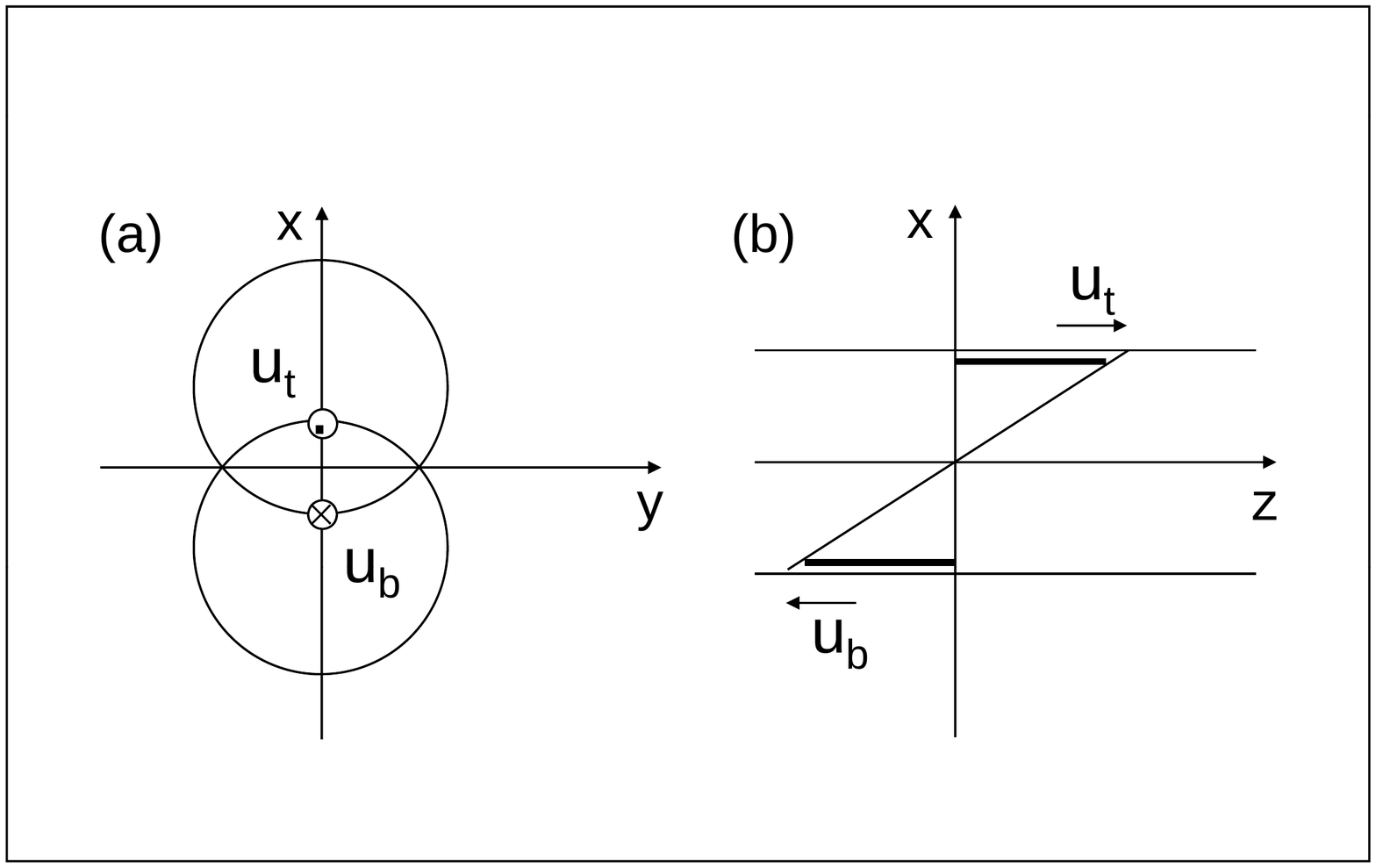}}
\vspace{-0.3cm}
\caption{
The sketch of a collision. Figure (a) is in the transverse, [x-y], plane
and (b) is in the reaction, [x-z] plane.
The almond shape in the middle of figure (a) is
the participant zone of the event. Right after the collision, streaks are
formed and the top streaks move along the $z$ direction while bottom ones
move along the $-z$ direction. This velocity shear may lead to the
Kelvin Helmholtz Instability, a wave formation, on the interface
between the top and bottom sheets.
}
\label{sketch}
\end{center}
\end{figure}

In idealized 2+1D model calculations the dissipation due to 3D expansion
is neglected, and thus the dissipation due to the 3D viscous expansion is
also neglected, which results in unrealistic estimates. The vorticity
of the flow is especially sensitive to such over-simplifications: while
in 2+1D the integrated vorticity is conserved in perfect fluid flow,
the decrease of vorticity is essential in realistic 3+1D CFD description.
\begin{figure}[ht]  
\begin{center}
\resizebox{0.95\columnwidth}{!}
{\includegraphics{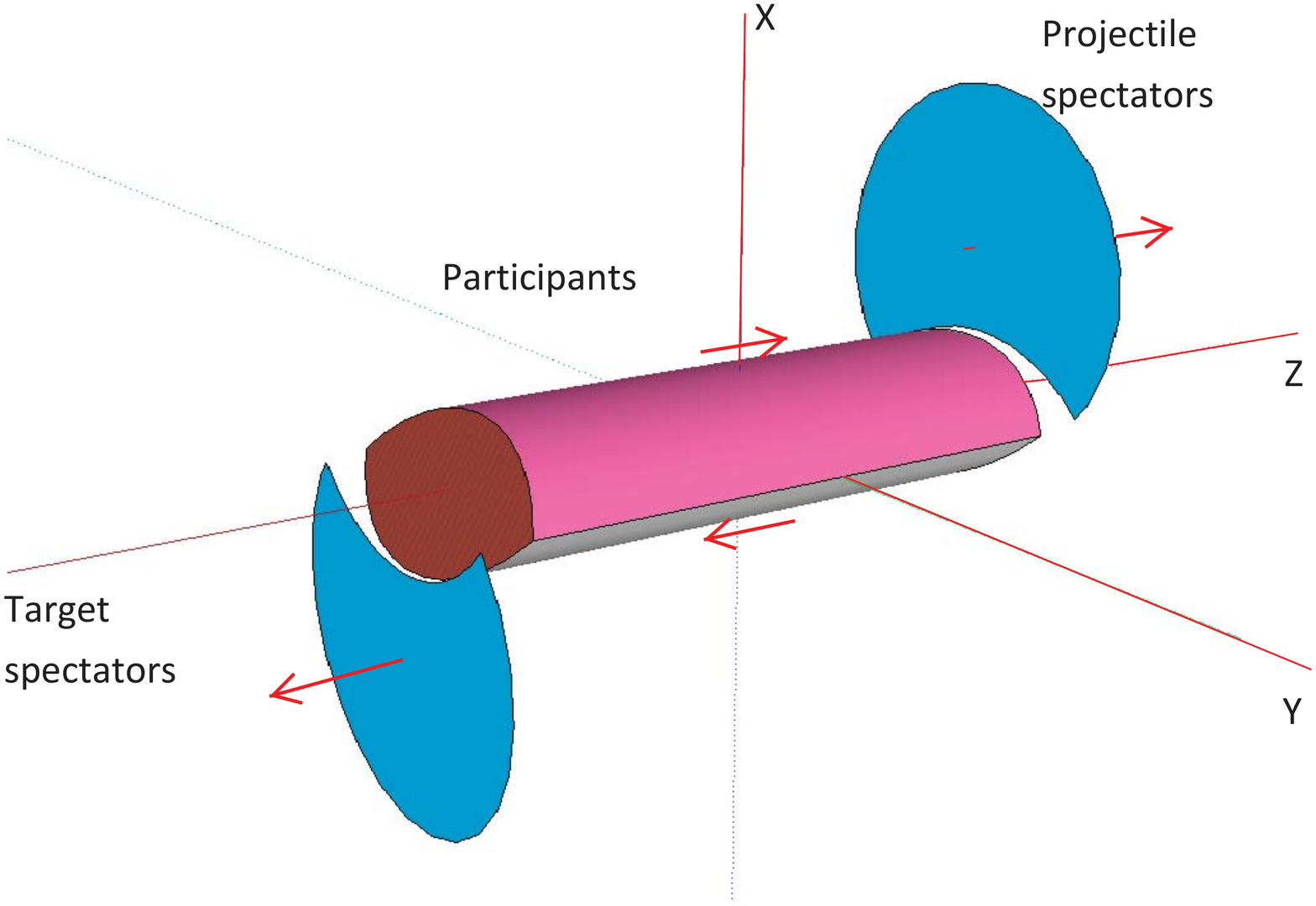}}
\vspace{-0.3cm}
\caption{
(Color online) The 3-dimensional view of the collision a few fm/c
after the impact corresponding to the situation illustrated in
Fig. \ref{sketch} (a).
The projectile spectators are going along the $z$-direction, the target
spectators are going along the $-z$ axis. We assume that the participants in
the middle, form a cylinder with an almond shape profile and
tilted end surfaces, where the top side is moving to the
right and the bottom is moving to the left. The participant cylinder
can be divided into streaks, each streak has its own velocity,
as shown in Fig. \ref{sketch} (b).
The velocity differences among the streaks result in the KHI effect.
}
\label{sketch2}
\end{center}
\end{figure}

In ref. \cite{Becattini} the angular momentum is assumed to have significant
effects on the longitudinal flow velocity and on its distribution
in the transverse plane, so that it gives rise to vorticity and
polarization. The arising polarization is also studied in
\cite{XNWang}, where a laminar shear flow is assumed where each layer has
different velocity, which is quite similar to the initial state velocity
profile depicted in Fig. \ref{sketch}. This type of initial state is
described in great detail in \cite{M2001-2}. In our present fluid
dynamical calculation we use this initial state model, which is tested
in several model calculations in the last decade. It describes correctly
the initial shear flow characteristics. The initial angular momentum is
based on the assumption that the initial angular momentum of the
participants (based on straight propagation geometry) is streak by streak
conserved. This leads to strong shear flow, shown in Fig. \ref{sketch2}.
This model assumes that the incoming Lorentz contracted nuclei
interpenetrate each other, and the leading charges in each streak
are slowed down by the large string-rope tension. This takes
about 3-5 fm/c for heavy nuclei depending on the impact parameter.
Then local equilibration is reached and the fluid dynamical evolution
starts.

In this work we study the development of vorticity in high energy
heavy ion reactions, and the development of the above mentioned specific
effects, which may arise in low viscosity \cite{Son,CKM} QGP.

We also compare the classical vorticity characteristics, and the effect
of the most dominant relativistic generalizations. These may provide
an insight into the possibilities of using these effects to precision
studies of the transport properties of QGP.

\begin{figure}[ht]  
\begin{center}
\resizebox{0.7\columnwidth}{!}
{\includegraphics{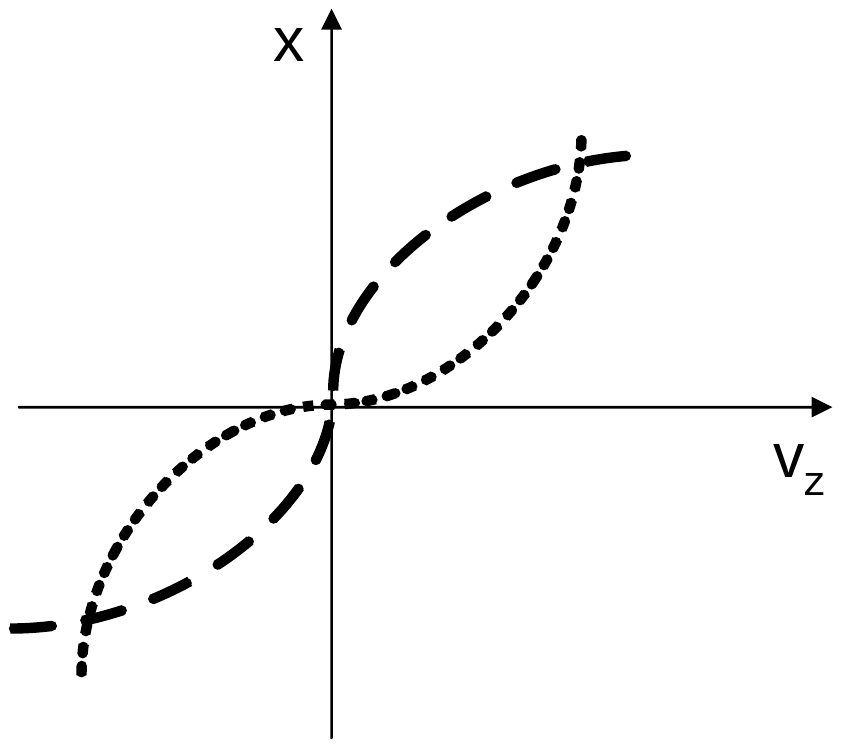}}
\caption{
The velocity along the $z$ axis, $v_z$, represented by the dotted curve
is calculated in our CFD model and presented in Fig.4 of Ref. \cite{hydro2}.
Here we have to mention that the velocity of the dotted curve will induce
the KHI effect, while the dashed curve will not, see chapter 8
of Ref. \cite{Drazin}.
}
\label{sketch3}
\end{center}
\end{figure}

It is also important to mention that the shear flow velocity profile
is essential from the point of instability. According to our initial
state description of the fluid dynamical calculations
(Fig. 4 of \cite{hydro2})  the velocity profile
along the $x$-axis is not linear but has a $x \sim \tan(v_z)$ shape
(dotted line in Fig. \ref{sketch3}), which may lead to KHI, in contrast
to a $x \sim \arctan(v_z)$ shape, which does not.

In the following we will discuss recent approaches to flow vorticity
in high energy heavy ion collisions, and present the vorticity
development and its distribution in the reaction plane.

In classical physics for incompressible, perfect fluids
vorticity exhibits an impressive conservation law: the conservation
of circulation. In high energy heavy ion physics the vorticity
definition must be modified, and the mentioned idealizations are
not applicable to energetic heavy ion reactions. Still, as CFD
calculations indicate typical flow patterns and instabilities
may occur here also. Thus their studies can provide insight
into the properties of the QGP fluid.

\begin{figure}[ht] 
\begin{center}
      \includegraphics[width=7.6cm]{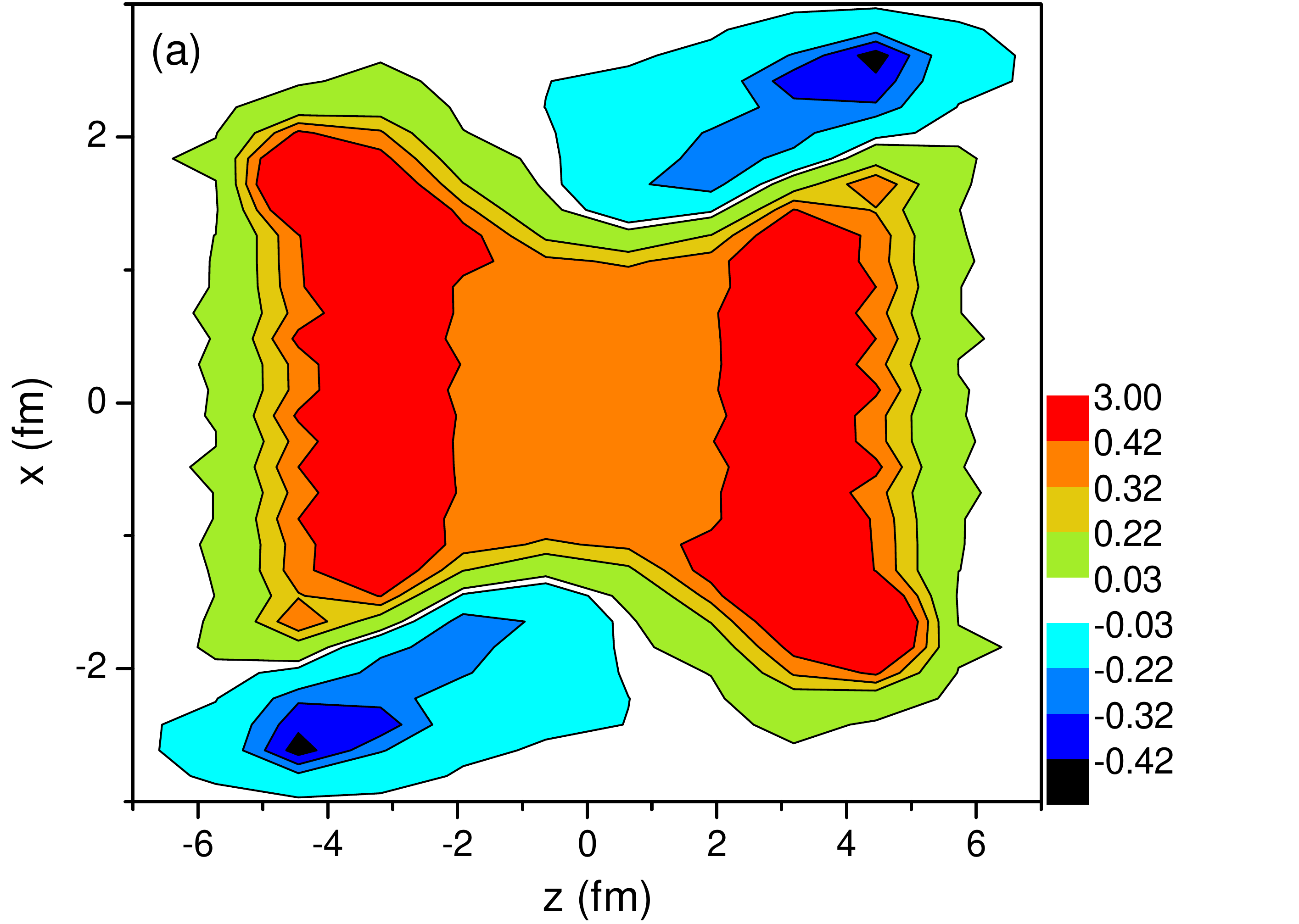}
      \includegraphics[width=7.6cm]{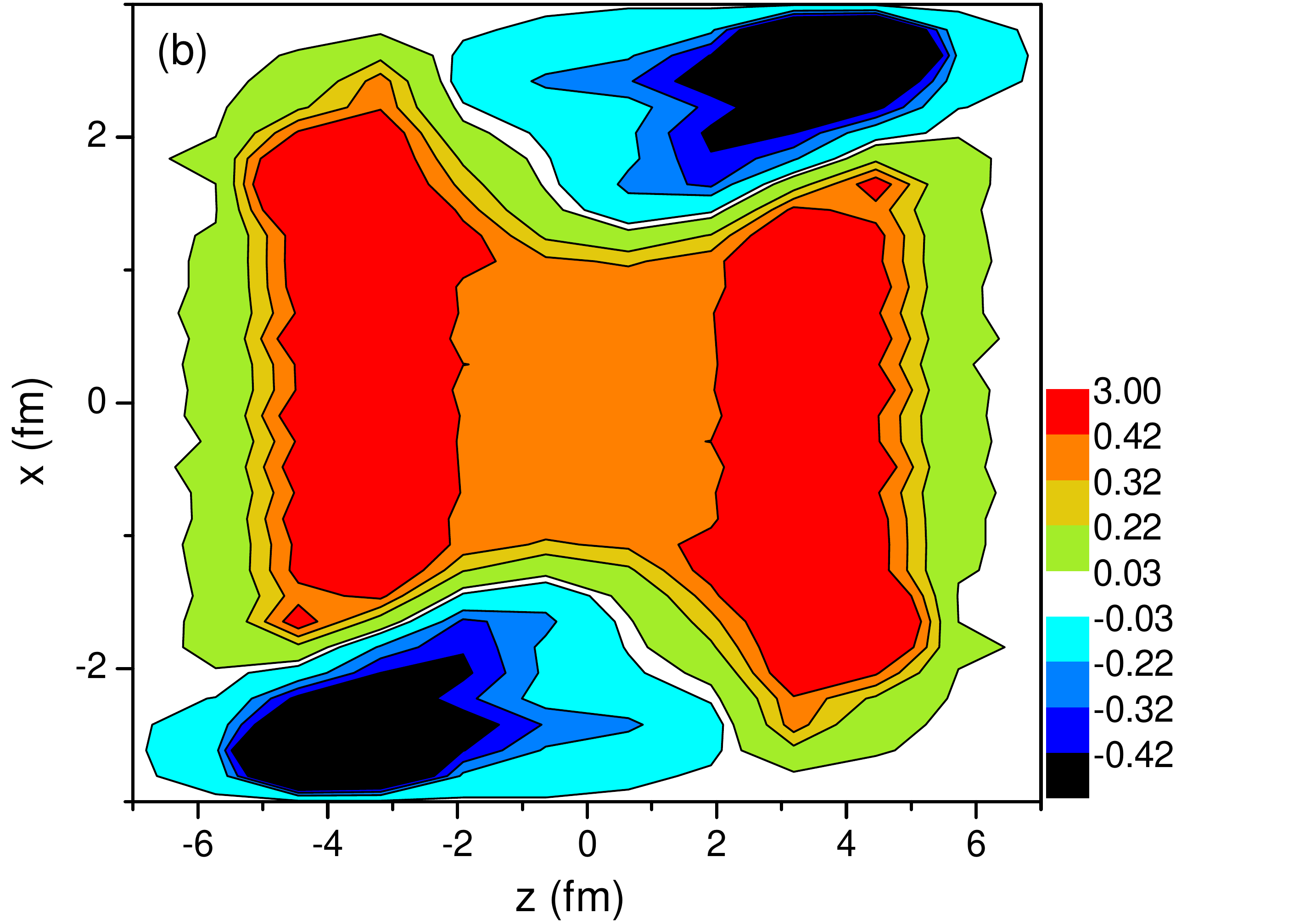}
\end{center}
\vspace{-0.3cm}
\caption{
(Color online)
The classical (a) and relativistic (b) weighted vorticity,
$\Omega_{zx}$ in units of c/fm, calculated in the reaction,
[x-z] plane at t=0.17 fm/c after the start of fluid dynamical evolution.
The collision energy is $\sqrt{s_{NN}}=2.76$ TeV and $b=0.7\,b_{max}$,
the cell size is $dx=dy=dz=0.4375$\,fm. The average vorticity in the
reaction plane is 0.1434 / 0.1185 c/fm for the classical / relativistic
weighted vorticities respectively.  }
\label{figG4}
\end{figure}

\section{The Vorticity}
\label{vorticity}

\paragraph{\bf Classically,}
%
in the reaction plane, [x-z], the vorticity  is defined as:
\be
\omega_y \ \equiv \ \omega_{xz} \ \equiv \ - \omega_{zx}
\equiv \ \frac{1}{2}(\partial_z v_x-\partial_x v_z)
\label{clvort}
\ee
where the $x, \ y, \ z$ components of the 3-velocity $\vec{v}$
are denoted by $v_x, \ v_y, \ v_z$ respectively.
In this definition we have already included the factor $\frac{1}{2}$ for the
symmetrization to have the same magnitude of vorticity as for symmetrized
volume divergence or expansion rate.

Here we study the
vorticity in the reaction plane (in the [x-z]-plane), at different
times in the CFD development.

In 3-dimensional space the vorticity can be defined as
\be
\vec{\omega} \equiv \frac{1}{2} {\bf rot}\ \vec{v}
=  \frac{1}{2} \, \nabla \times \vec{v} \,
\label{wc}
\ee
(but this cannot be generalized to four or more dimensions).
The {\em circulation} of the flow is the integral of the velocity along
a closed curve, $C$, with the line element  $d\vec{l}$. It is defined as
$$
\Gamma = \oint_C \vec{v}\, d\vec{l} = \int_A 2 \vec{\omega}\, d\vec{A} \ ,
$$
where $A$ is an (arbitrary) surface surrounded by the curve $C$,
and the normal of its surface element is d$\vec{A}$.
\begin{figure}[ht] 
\begin{center}
      \includegraphics[width=7.6cm]{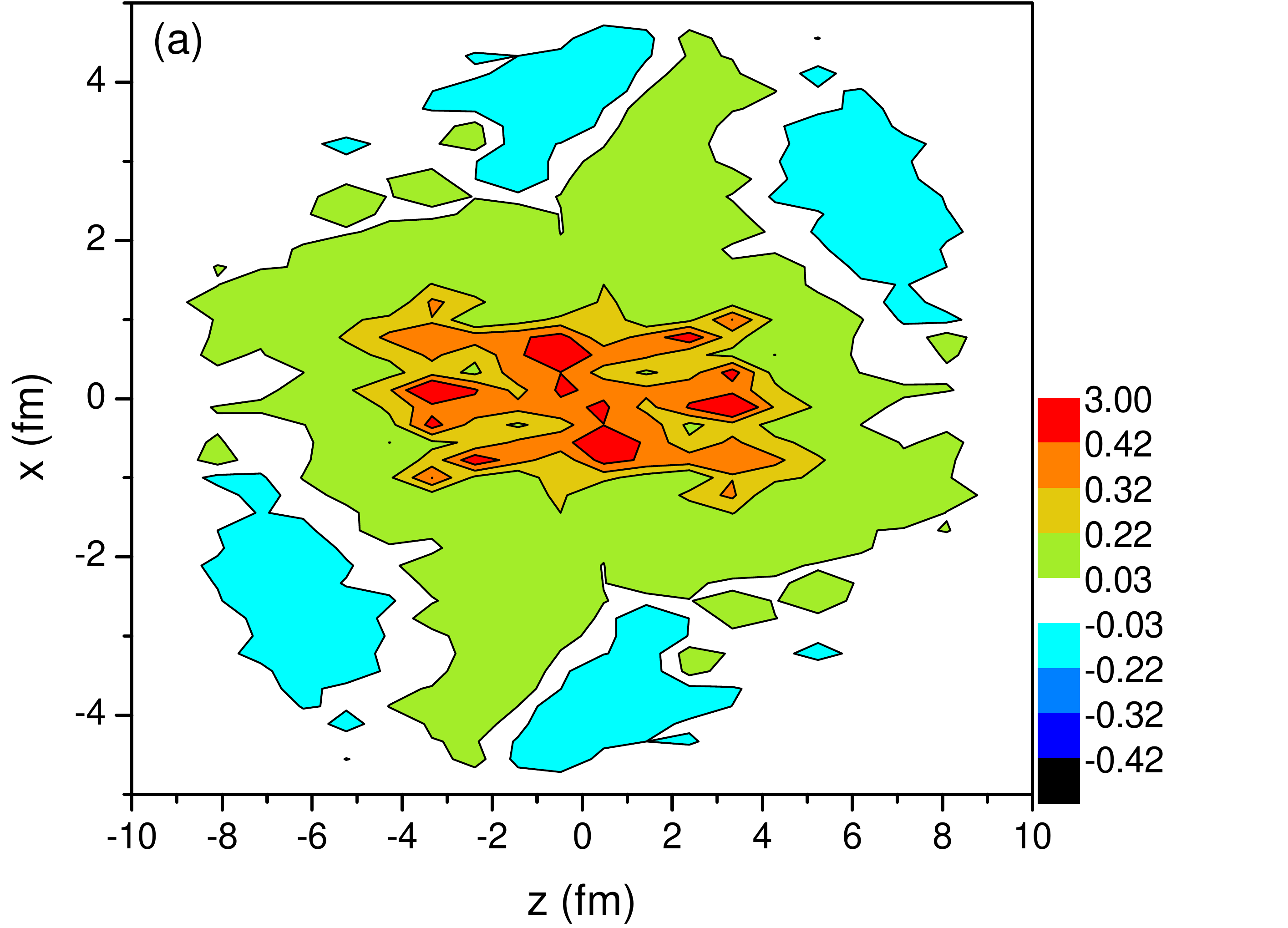}
      \includegraphics[width=7.6cm]{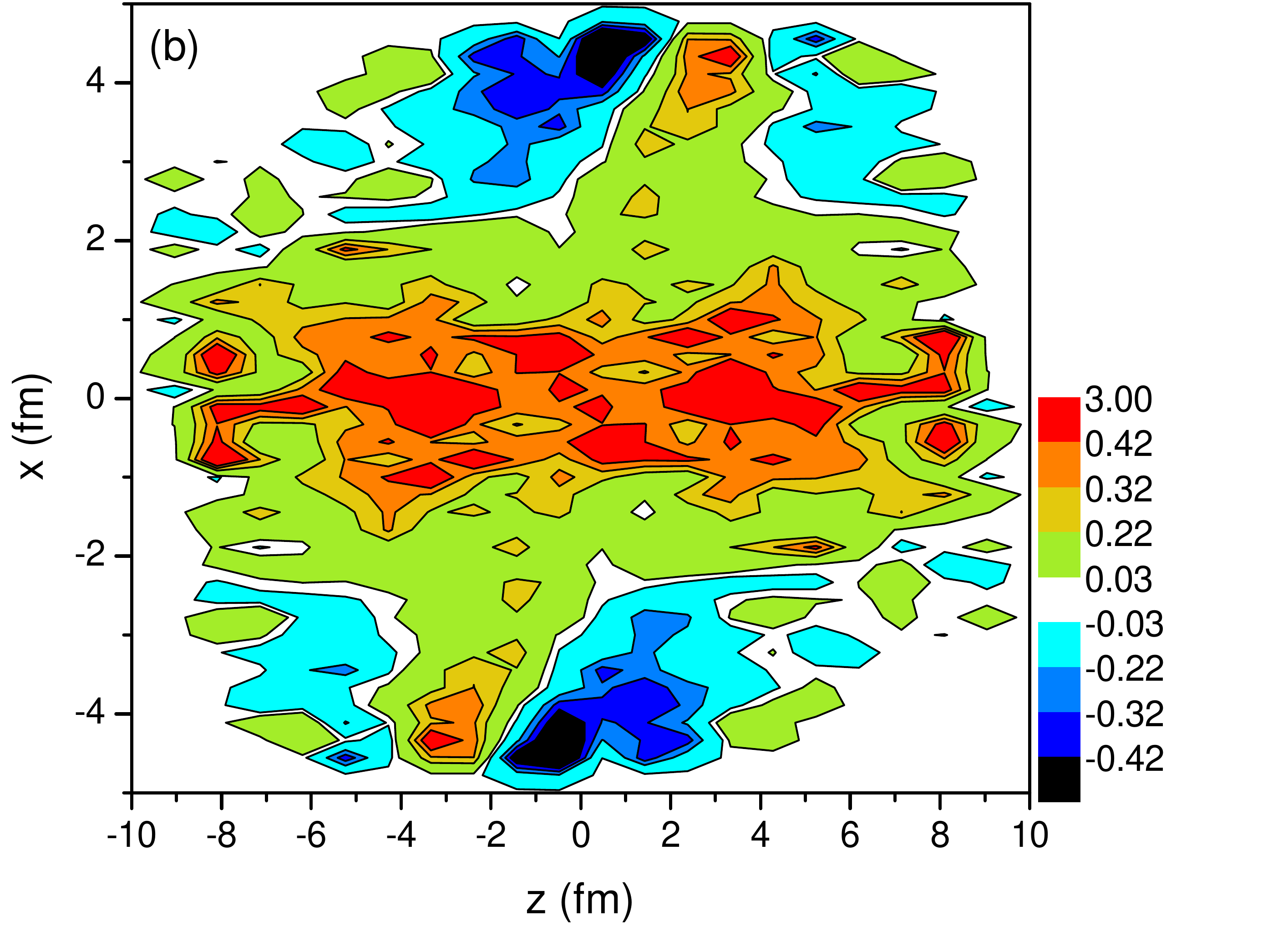}
\end{center}
\vspace{-0.3cm}
\caption{
(Color online)
The classical (a) and relativistic (b) weighted vorticity,
$\Omega_{zx}$, (c/fm), calculated in the reaction [x-z] plane at t=3.56 fm/c.
The collision energy is $\sqrt{s_{NN}}=2.76$ TeV and $b=0.7\,b_{max}$,
the cell size is $dx=dy=dz=0.4375 fm$. The average vorticity in the
reaction plane is 0.04845 / 0.07937 c/fm for the classical / relativistic
weighted vorticity respectively.}
\label{figG84}
\end{figure}
For rotationless (potential) flow the circulation vanishes. In peripheral
heavy ion reactions the flow is rotational. In classical fluid
dynamics the circulation is constant in a fluid along the
line of motion of a fluid element if the viscosity of the flow vanishes and
flow is barotropic, i.e.
the pressure depends only on the density of the fluid, $P = P(\rho)$.
These conditions are not satisfied for QGP in heavy ion reactions.

In the relativistic case the formulation of fluid dynamics is
more involved, and the formalism must be modified \cite{Molnar,Stefan}.
The three- velocities are replaced by four- velocities, the derivatives
should take into account the world lines of the particles and
the changes of these with time and the mass density is replaced
by the energy density where the pressure will have a non-negligible
role. This last modification e.g. leads to a modified relativistic
definition of the vorticity and circulation, which includes
the specific enthalpy of the fluid \cite{Florkowski2010}.
This would then extend the
validity of the conservation of circulation under the same conditions.
On the other hand this modified relativistic vorticity would then
have a different dimension and it would not be conserved anyway in
heavy ion reactions where the conditions of validity are not satisfied.

\begin{figure}[ht] 
\begin{center}
      \includegraphics[width=7.6cm]{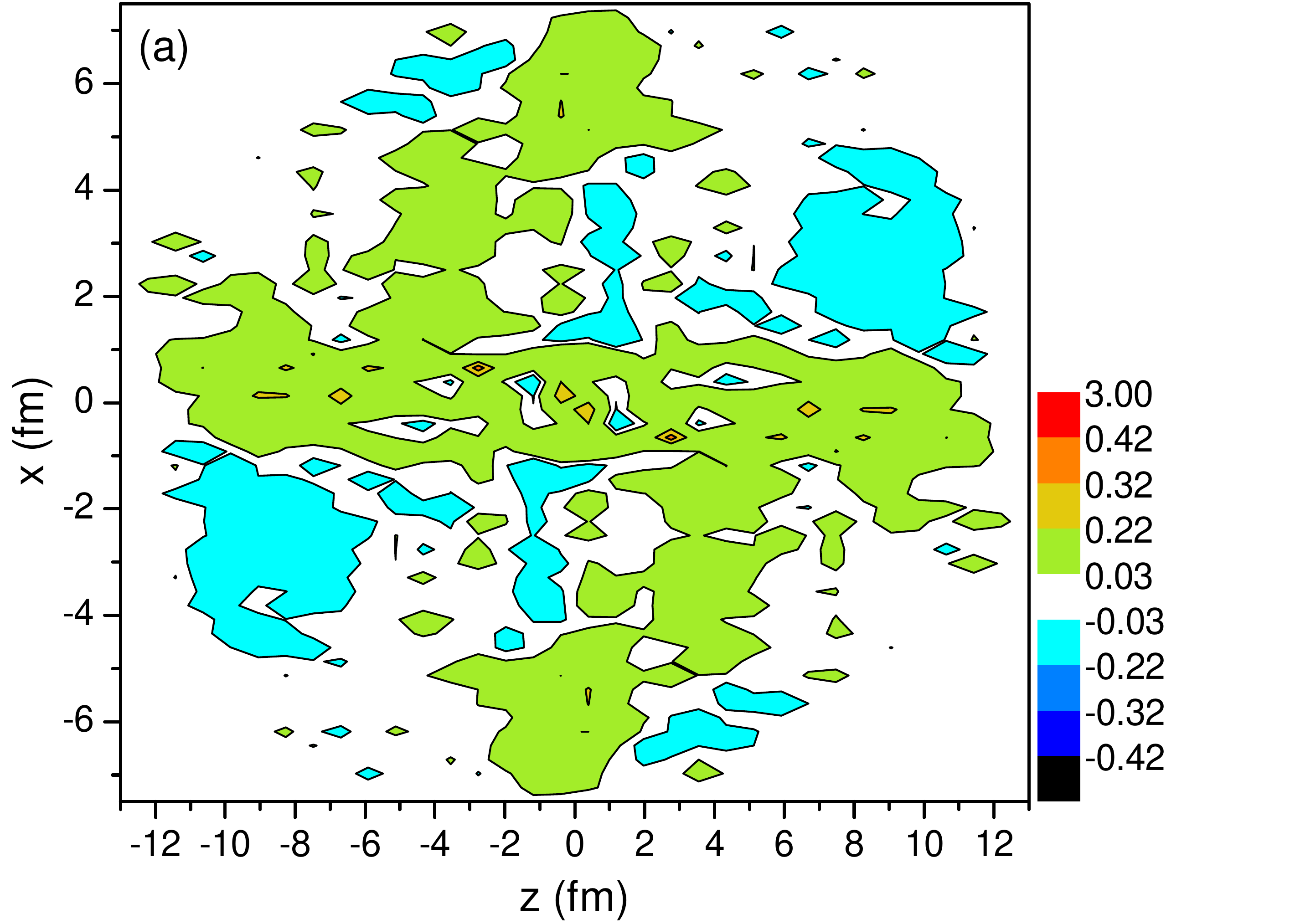}
      \includegraphics[width=7.6cm]{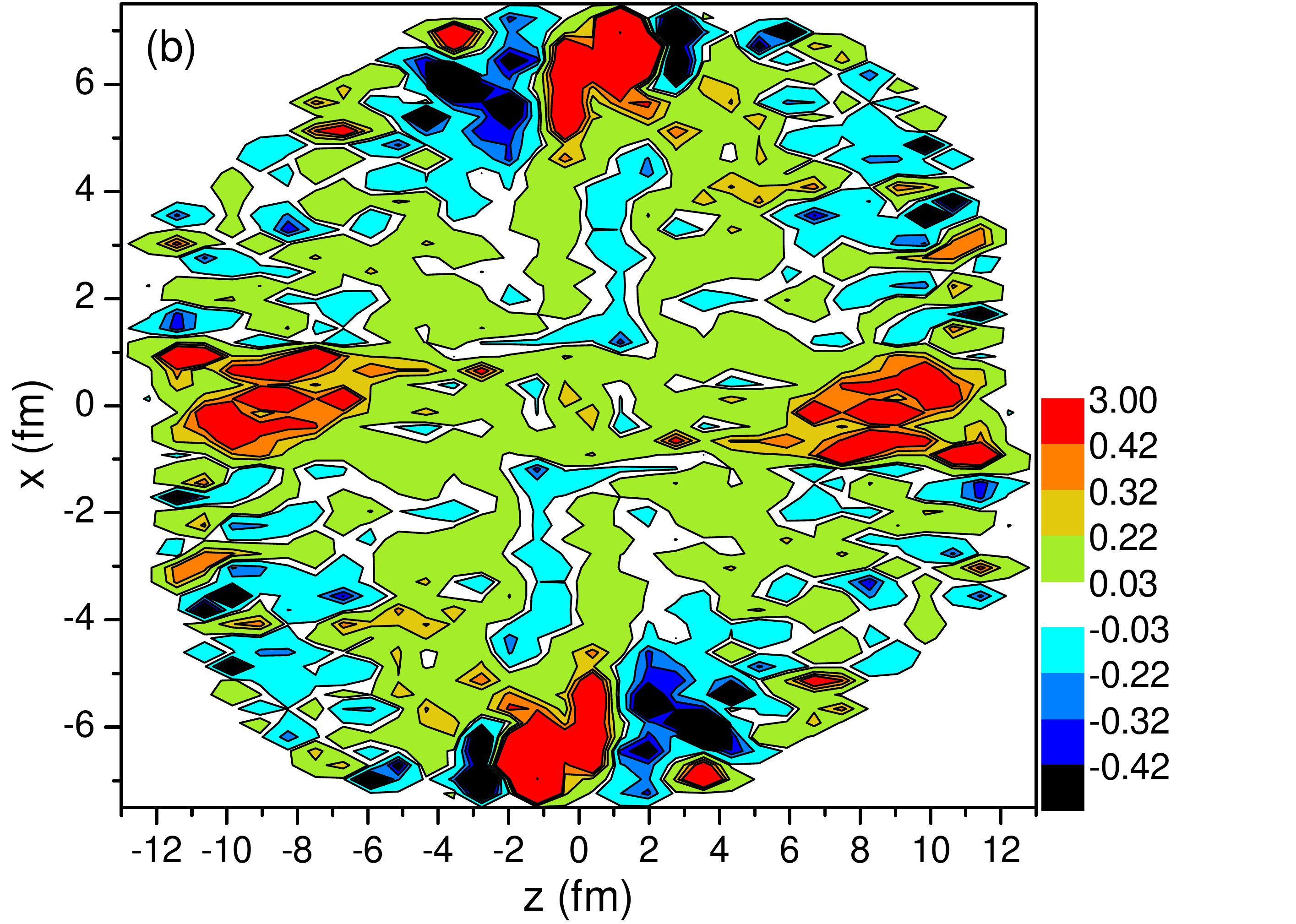}
\end{center}
\vspace{-0.3cm}
\caption{
(Color online)
The classical (a) and relativistic (b) weighted vorticity
$\Omega_{zx}$,(c/fm), calculated in the reaction [x-z] plane at t=6.94 fm/c.
The collision energy is $\sqrt{s_{NN}}=2.76$ TeV and $b=0.7\,b_{max}$,
the cell size is $dx=dy=dz=0.4375 fm$ The average vorticity in the
reaction plane is 0.01555 / 0.05881 c/fm for the classical / relativistic
weighted vorticity respectively.}
\label{figG164}
\end{figure}

Thus, in the present work we intend to keep the possibility of
comparison to the classical vorticity and circulation, in a relativistic
system where the (baryon charge) density and temperature change
violently, the pressure is not negligible and does not depend on the
density only. Still the energy and momentum, ($T^{00}$ and $T^{i0}$)
are strictly conserved, the total energy remains constant while the
momentum vanishes (in the C.M. frame).

Keeping in view fluid dynamics with fluid elements, we introduce
a weight proportional to the energy content of the
fluid element.
The energy distribution (and thus the local angular momentum) of the
matter is highly non-homogeneous in a heavy ion reaction.
To reflect our physical situation better, we weight the contribution
of our fluid cells by the local energy density in the
reaction plane, however, without changing the average vorticity
in the layer of the reaction plane. Thus we weight with a distribution
normalized to unity.
We define an energy density weighted, average vorticity as
\be
\Omega_{zx} \equiv \  w(z,x) \ \ \omega_{zx}
\ee
so that this weighting does not change the average circulation
of the layer, i.e.,
the sum of the average of the weights over all fluid cells is unity,
$\langle w(z,x) \rangle \ = \ 1$,
both if we consider one y-layer only, or if we consider all y-layers.
This weighting does not change the average vorticity value of the
set, just the cells will have larger weight with more energy content,
$T^{00}$.

\begin{figure}[ht] 
\begin{center}
      \includegraphics[width=7.6cm]{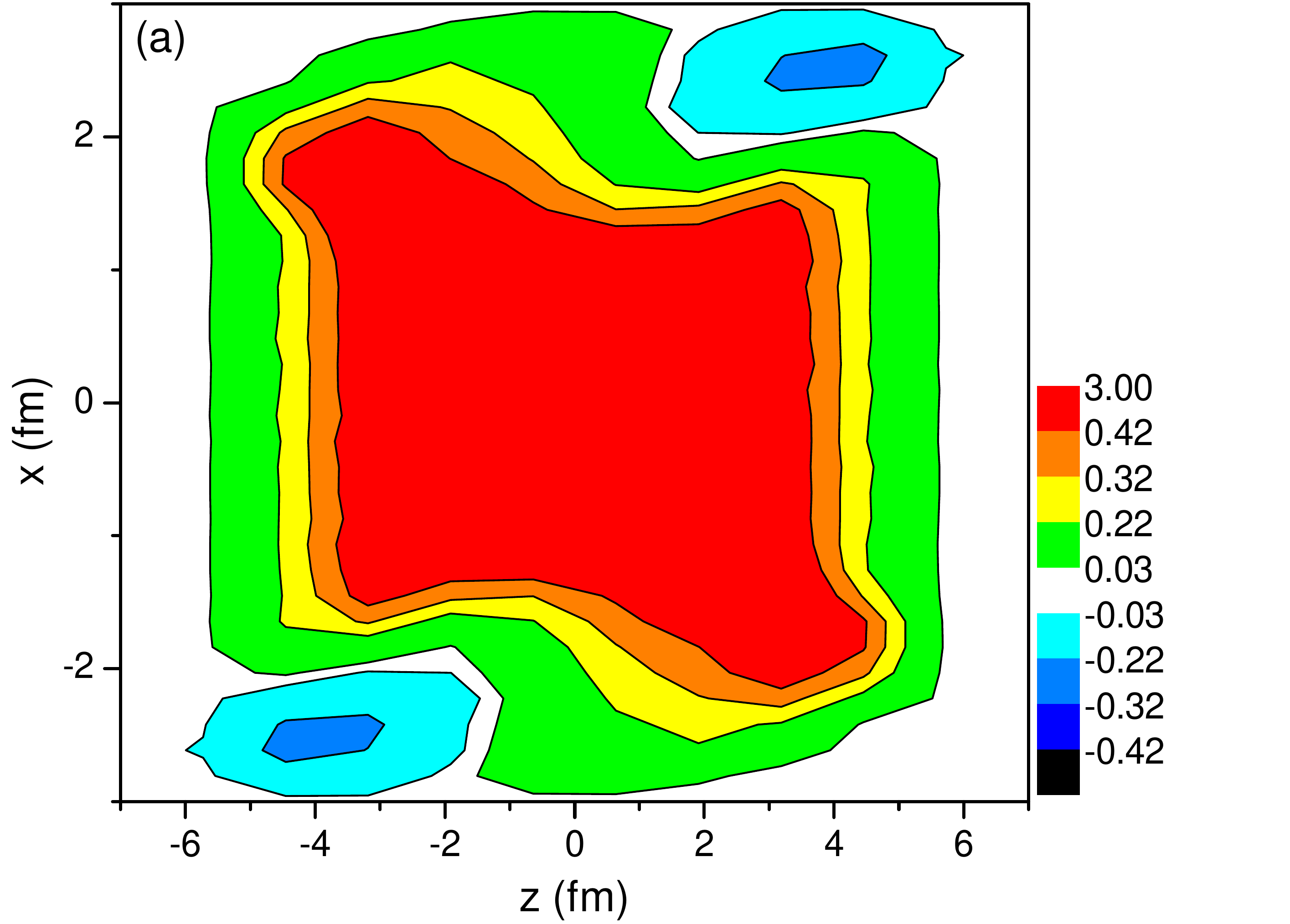}
      \includegraphics[width=7.6cm]{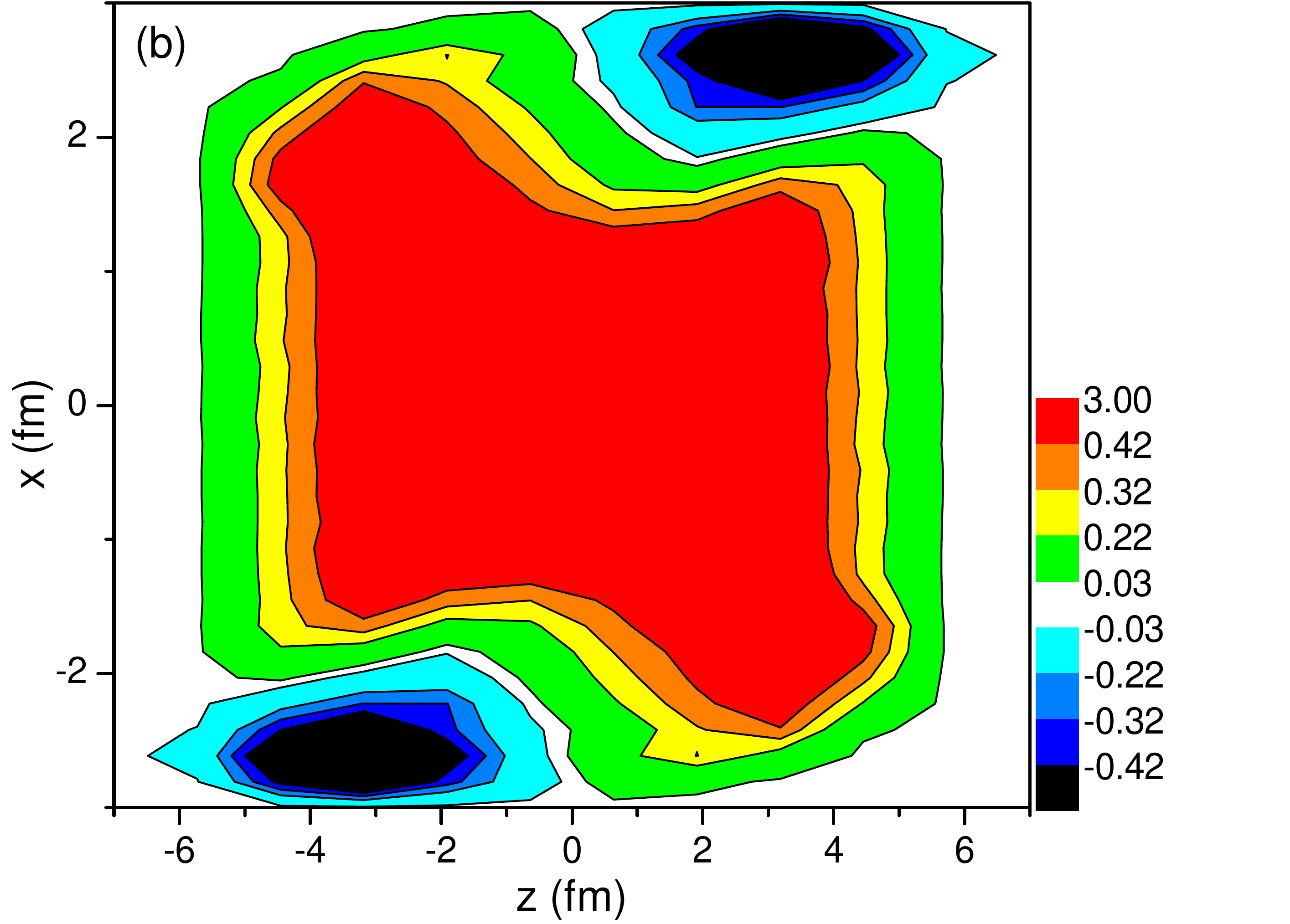}
\end{center}
\vspace{-0.3cm}
\caption{
(Color online)
The classical (a) and relativistic (b) weighted vorticity
calculated for all [x-z] layers at t=0.17 fm/c.
The collision energy is $\sqrt{s_{NN}}=2.76$ TeV and $b=0.7\,b_{max}$,
the cell size is $dx=dy=dz=0.4375$ fm. The average vorticity in the
reaction plane is 0.1971 / 0.19004 c/fm for the classical / relativistic
weighted vorticity respectively. }
\label{figG4-al}
\end{figure}

As we have discretized fluid cells, we study separately the reaction plane,
one y-layer, or the whole system, all y-layers.
The total energy content of a cell at point $(z,x)$ or
the corresponding $i,k$, is $E_{ik}=T^{00}(z,x)$. The total energy
in a y-layer (or in all y-layers)
is $E_{tot} = \sum_{ik} E_{ik}$, while the number of the cells is
in a y-layer (or in all y-layers)
is $N_{cell}$. Thus the average energy for for a fluid cell is
$E_{tot}/N_{cell}$ in both cases. We divide the actual energy of a
fluid cell, $E_{ik}$, with the average fluid cell energy
$E_{tot}/N_{cell}$, so that the vorticity values on average
will remain comparable with the non-weighted values, but still
larger energy cells will have more weight. The total energy
$\sum_{All\, cells} T^{00}$ remains exactly constant in our
case (in the numerical calculation to $10^{-16}$ accuracy).
Of course this is not true for a single layer like the reaction plane.

Our weight then, should be proportional with the local energy density
$$
w_{ik} \equiv \frac{E_{ik}}{(E_{tot}/N_{cell})} \ .
$$

\begin{figure}[ht] 
\begin{center}
      \includegraphics[width=7.6cm]{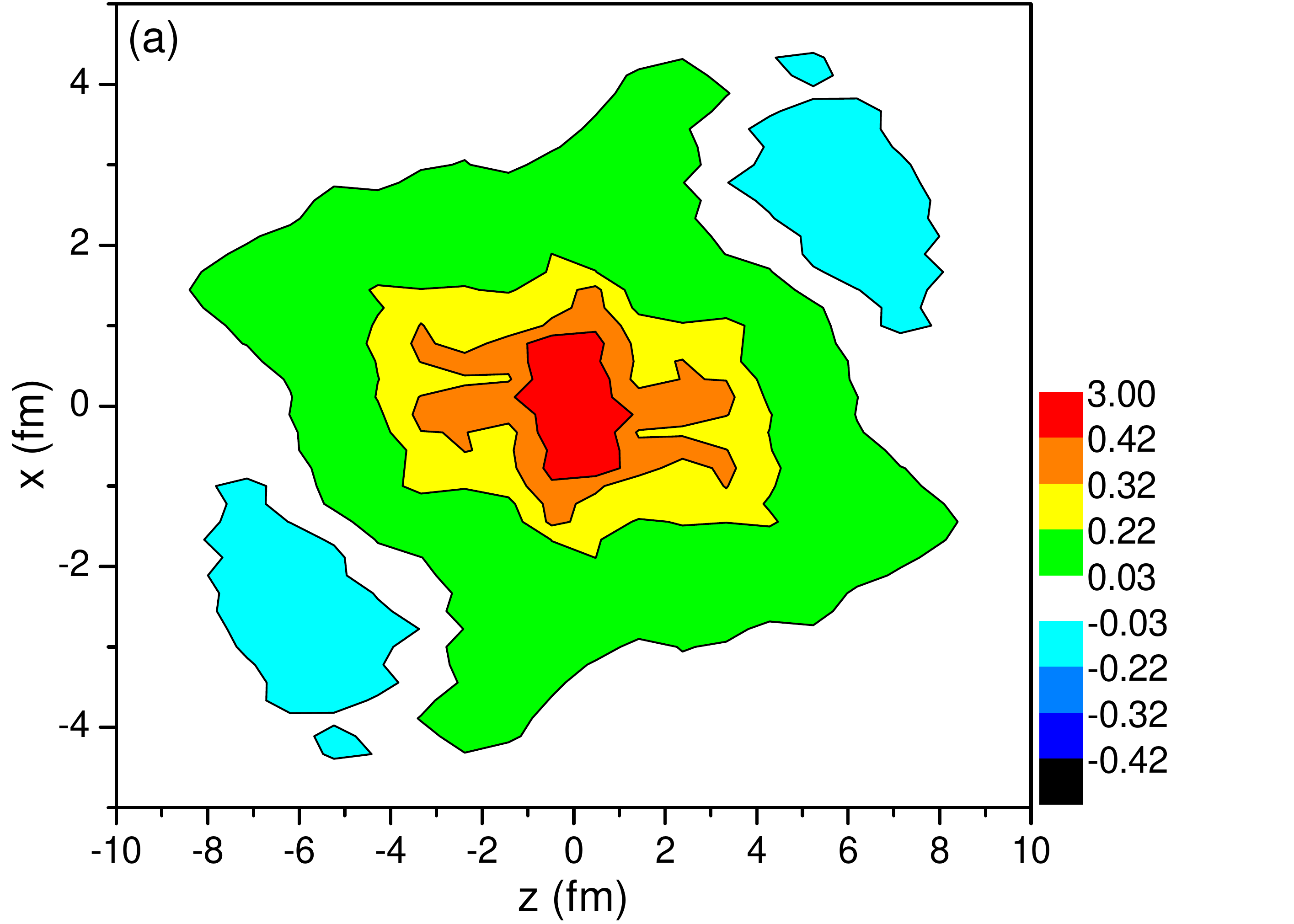}
      \includegraphics[width=7.6cm]{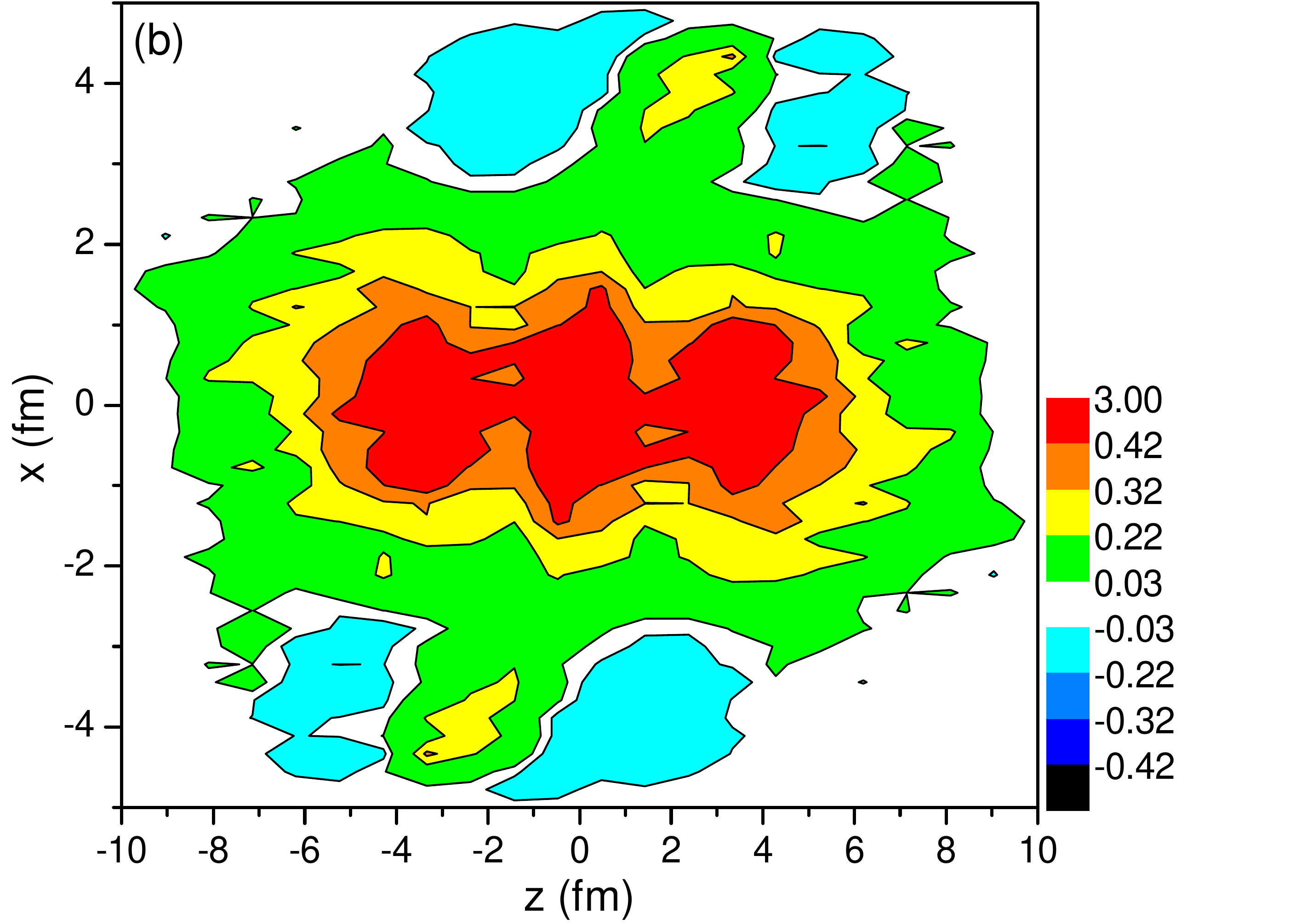}
\end{center}
\vspace{-0.3cm}
\caption{
(Color online)
The classical (a) and relativistic (b) weighted vorticity
calculated for all [x-z] layers at t=3.56 fm/c.
The collision energy is $\sqrt{s_{NN}}=2.76$ TeV and $b=0.7\,b_{max}$,
the cell size is $dx=dy=dz=0.4375$ fm.  The average vorticity in the
reaction plane is 0.0538 / 0.10685 c/fm for the classical / relativistic
weighted vorticity respectively. }
\label{figG84-al}
\end{figure}

Within the reaction plane, at a given moment of time, the cells carrying
larger amount of energy will get a larger weight than those, which
carry less energy.
The edge cells, carrying less energy, show stronger fluctuations.

This average vorticity weighted by the cell energy, which is a conserved
quantity for all cells, provides a possibility to compare the
results to classical systems and their features.

An alternative method to present the relativistic vorticity
is presented in ref. \cite{Florkowski2010} by weighting with
the specific entropy, would provide conserved circulation, $\Gamma$,
if the pressure were exclusively density dependent.
As it was mentioned earlier, this is not
the case for quark-gluon plasma, so for us this advantage is
not realized, while this weighting changes both the value and
the dimension of the vorticity, so it would make the comparison
to classical results difficult.

The weighted vorticity in the reaction, [x-z] plane at different time
steps are shown in the Figs. \ref{figG4}, \ref{figG84}, \ref{figG164}.

\begin{figure}[ht] 
\begin{center}
      \includegraphics[width=7.6cm]{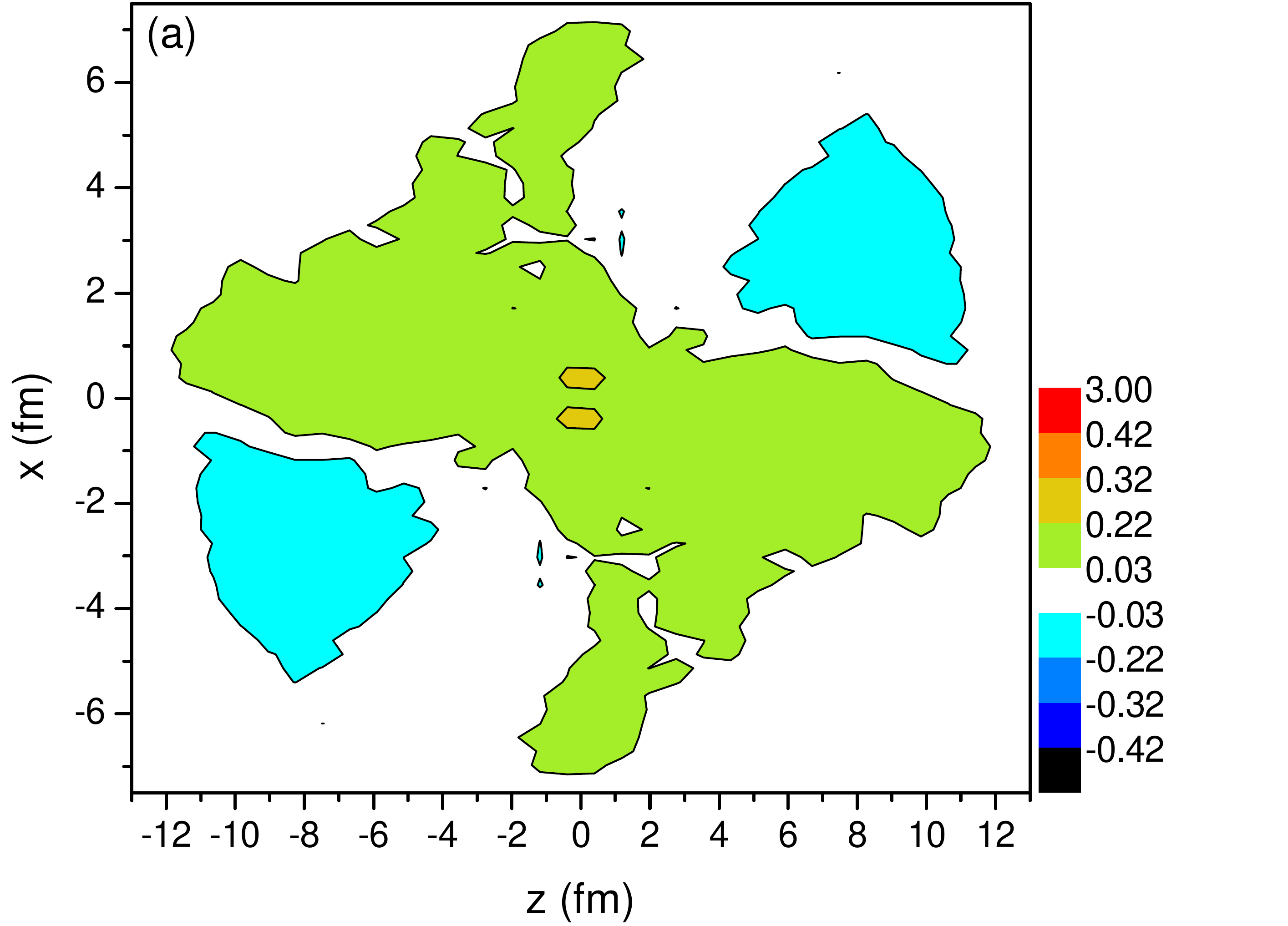}
      \includegraphics[width=7.6cm]{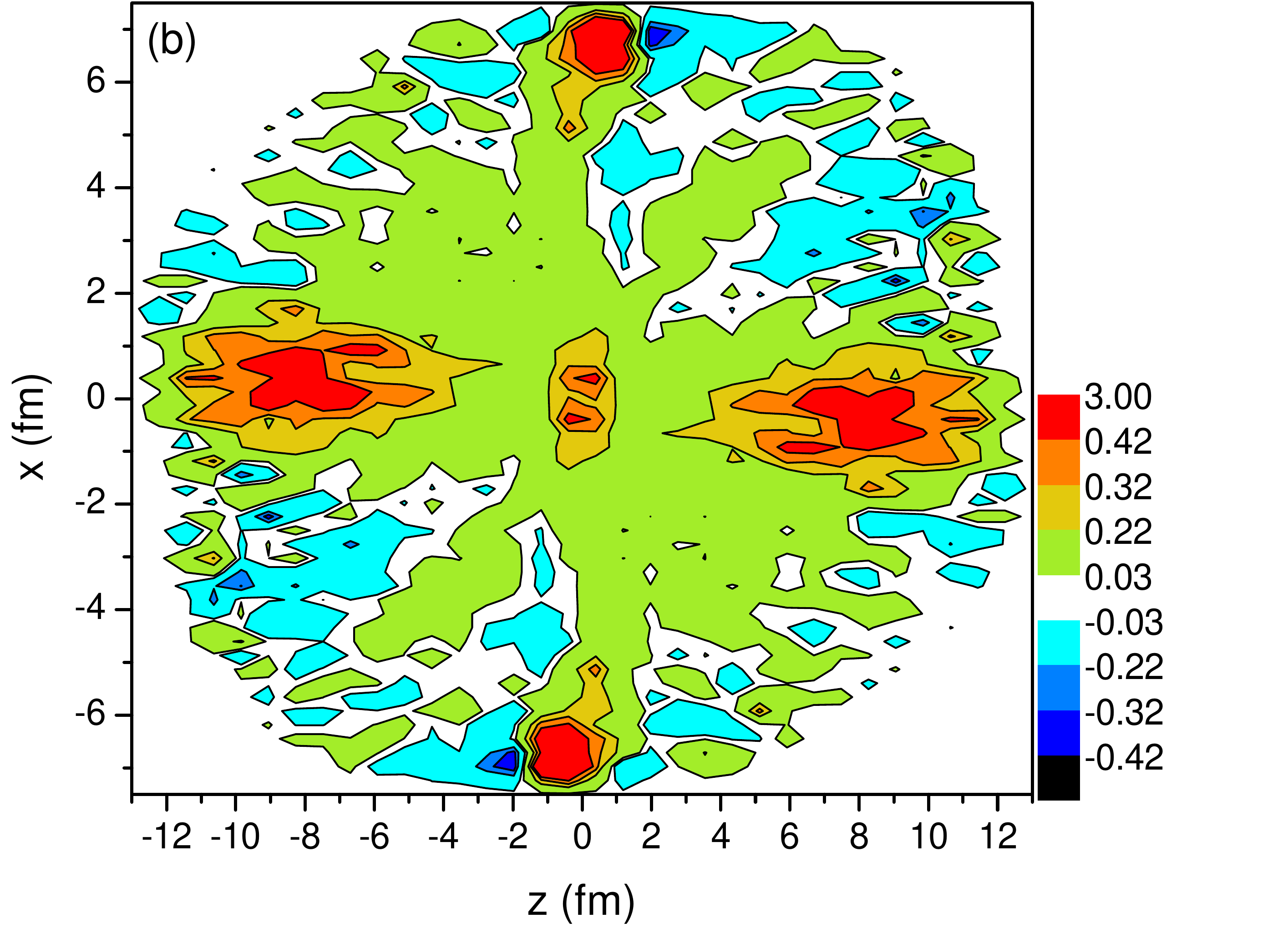}
\end{center}
\vspace{-0.3cm}
\caption{
(Color online)
The classical (a) and relativistic (b) weighted vorticity
calculated for all [x-z] layers at t=6.94 fm/c.
The collision energy is $\sqrt{s_{NN}}=2.76$ TeV and $b=0.7\,b_{max}$,
the cell size is $dx=dy=dz=0.4375$ fm.  The average vorticity in the
reaction plane is 0.0159 / 0.05881 c/fm for the classical / relativistic
weighted vorticity respectively.}
\label{figG164-al}
\end{figure}

\paragraph{\bf In relativistic flow,}
we are following the definition of in Ref. \cite{Molnar},
for the relativistic case.
The expansion rate, $\Theta$ and the vorticity tensor,
$\omega^\mu_\nu$ are defined as
\be
\Theta \equiv \nabla_\mu u^\mu = \partial_\mu u^\mu ,
\ee
\be
\omega^\mu_\nu \equiv \frac{1}{2} (\nabla_\nu u^\mu - \nabla^\mu u_\nu) ,
\ee
were for any four vector $q^\mu$ the quantity
$\nabla_\alpha q^\mu \equiv \Delta^\beta_\alpha \, \partial_\beta q^\mu =
\Delta^\beta_\alpha \, q^\mu_{,\beta}$  and
$\Delta^{\mu\nu} \equiv g^{\mu\nu} - u^\mu u^\nu$.
This leads to
\begin{eqnarray*}
\omega^\mu_\nu &=&
\frac{1}{2} \Delta^{\mu\alpha} \Delta^\beta_\nu
(u_{\alpha,\beta} - u_{\beta,\alpha})
\nonumber \\
&=& \frac{1}{2}
\left[ (\partial_\nu u^\mu - \partial^\mu u_\nu) +
       (u^\mu u^\alpha \partial_\alpha u_\nu -
        u_\nu u^\alpha \partial_\alpha u^\mu) \right]
\nonumber \\
&=& \frac{1}{2}
\left[ (\partial_\nu u^\mu - \partial^\mu u_\nu) +
       (u^\mu \partial_\tau u_\nu -
        u_\nu \partial_\tau u^\mu) \right] ,
\end{eqnarray*}
where
$\partial_\tau u^\mu \equiv \dot{u}^\mu= u^\alpha \partial_\alpha u^\mu$
is the proper time derivative of $u^\mu$.

In our initial state in the middle part of
the collision the streaks are stopped after two Lorentz contracted ions
collide and interpenatrate each other. If the acceleration of the
fluid elements is negligible compared to the rotation,
$|\partial_{\tau}u^{\mu}|\ll|\partial_x u^z|$, this assumption holds for
the initial moments in our model, which has strong initial shear flow.
Thus the second term can be dropped. This is also the case considered in
Ref.\cite{Stefan}, while this work is studying the vorticity in the
[x-y] plane instead of the reaction plane. In this case  the relativistic
vorticity is:
\be
\omega_{\mu \nu}=\frac{1}{2}(\partial_{\nu}u_{\mu}-\partial_{\mu}u_{\nu})
\label{wuv}
\ee
where
\begin{eqnarray*}
\partial^{\nu} &=&(\partial_0,\partial_x,\partial_y,\partial_z)\\
u_{\mu}&=&\gamma(1,-v_x,-v_y,-v_z)
\end{eqnarray*}

Let us expand Eq. (\ref{wuv}) in four dimensions:

\ba
\omega_{\;\mu}^\nu&=&
\frac{1}{2}
\left[
\left(
  \begin{array}{cccc}
    \partial_0 \gamma  &  -\partial_0 \gamma v_x  &  -\partial_0 \gamma v_y  &  -\partial_0 \gamma v_z\\
    \partial_x \gamma  &  -\partial_x \gamma v_x  &  -\partial_x \gamma v_y  &  -\partial_x \gamma v_z\\
    \partial_y \gamma  &  -\partial_y \gamma v_x  &  -\partial_y \gamma v_y  &  -\partial_y \gamma v_z\\
    \partial_z \gamma  &  -\partial_z \gamma v_x  &  -\partial_z \gamma v_y  &  -\partial_z \gamma v_z\\
  \end{array}
\right)
\right.
\nonumber \\
&-&
\left.
\left(
  \begin{array}{cccc}
     \partial_0 \gamma      &       \partial_x \gamma  &       \partial_y \gamma  &    \partial_z \gamma   \\
    -\partial_0 \gamma v_x  &  -\partial_x \gamma v_x  &  -\partial_y \gamma v_x  &  -\partial_z \gamma v_x\\
    -\partial_0 \gamma v_y  &  -\partial_x \gamma v_y  &  -\partial_y \gamma v_y  &  -\partial_z \gamma v_y\\
   - \partial_0 \gamma v_z  &  -\partial_x \gamma v_z  &  -\partial_y \gamma v_z  &  -\partial_z \gamma v_z\\
  \end{array}
\right)
\right]
\ea
and we see that the vorticity is an antisymmetric tensor.
Here for the vorticity development in the reaction plane
we calculate $\omega_{24}$:
\ba
\omega^x_{\;z} &=& - \omega^z_{\;x}
\nonumber \\
&=&
 \frac{1}{2}(\partial_z\gamma v_x-\partial_x\gamma v_z)
\nonumber \\
&=&
\frac{1}{2}\gamma (\partial_z v_x-\partial_x v_z)
+\frac{1}{2}(v_x\partial_z \gamma-v_z\partial_x\gamma)
\label{wt}
\ea

The fluid cells were weighted the same way as in the non-relativistic
vorticity estimate. Due to the fact that the relativistic vorticity
includes the relativistic $\gamma$ factor, eq. (\ref{wt}), the vorticity
increases, especially at the edges where the cells have larger
flow velocities.

\begin{figure*}[ht] 
\begin{center}
      \includegraphics[width=12cm]{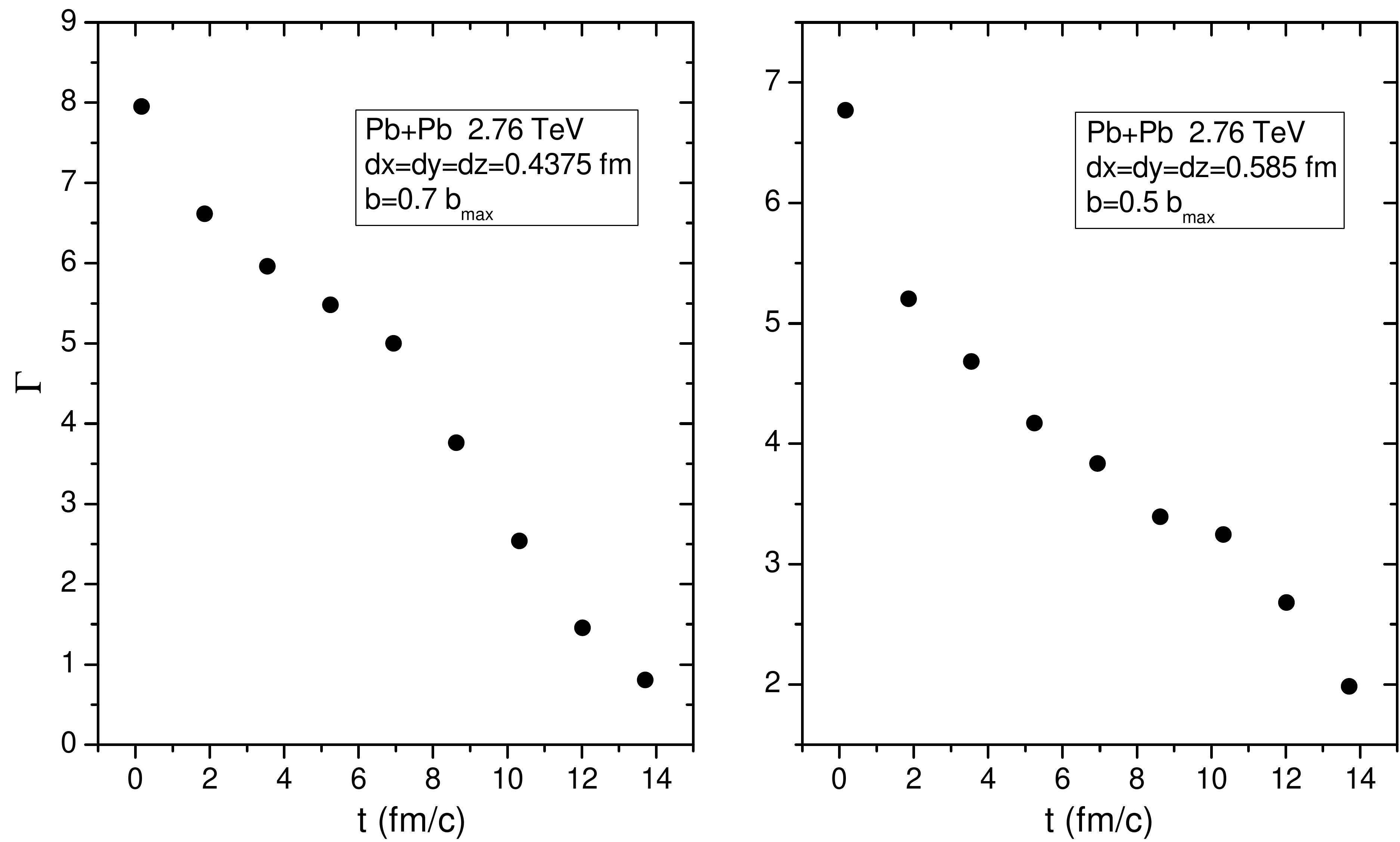}
\end{center}
\vspace{-0.3cm}
\caption{
The time dependence of classical circulation, $\Gamma(t) $,
in units of [fm\,c],
calculated for all [x-z] layers and then taking the average of the
circulations for all layers.
The collision energy is $\sqrt{s_{NN}}=2.76$ TeV and $b=0.7\,b_{max}$,
the cell size is $dx=dy=dz=0.4375$ fm (left). For comparison another
initial state configuration was also tested for the same collision
energy but  $b=0.5\,b_{max}$,
the cell size is $dx=dy=dz=0.585$ fm (right). This configuration
shows also the rotation, but due to its less favorable parameters
it does not show the KHI. Although at this impact parameter,
which is less peripheral the reaction plane has a larger
area filled with matter, nevertheless the initial classical circulation
is less by about 15\%. For the more peripheral case with
smaller numerical viscosity
the circulation decreases with time faster and the circulation
for the two cases becomes equal around $t = 8-10 \,$ fm/c.}
\label{Gamma-cl}
\end{figure*}

The weighted relativistic vorticity distributions in the reaction,
[x-z] plane at different time  steps are shown in the
Figs. \ref{figG4}, \ref{figG84}, \ref{figG164}.
At the last time step presented, in the reaction plane we have already
an extended area occupied with matter. In case of peripheral
reactions the multiplicity is already small, thus the fluctuations in
the reaction plane are considerable. In the relativistic
case the outside edges show larger vorticity and
the random fluctuations are still strong.

\section{Weighted Vorticity Distributions}

The relativistic vorticity distributions have increased
amplitudes compared to the classic ones, due to the relativistic
$\gamma$-factor and its derivatives in the relativistic expression.
This is especially visible at the edges where the flow velocities are
the largest.

The amplitude of weighted vorticity decreases with time as the
matter expands. Random fluctuations are apparent at late times
for the dilute matter, especially at the edges, and particularly for
the relativistic vorticity, which has enhanced amplitudes.

With increasing time the fluid expands, and outside more dense
shell develops with a less dense central zone. This feature is
also apparent in the vorticity distribution of the mater
in the central, reaction plane, where the central part has
smaller weighted vorticity amplitude.

The non-central, parallel layers, at increasing y-values,
have similar positive weighted vortices, $\Omega_{zx}^{y_i}$.
These outside layers have narrower boundaries in the $x$-direction
and these fall into the denser outside zone of the expanding matter.

We can average the weighted vorticity distributions over all
layers parallel to the reaction plane.  This compensates the
low density central zone in the reaction plane and
leads to a more uniform, layer-averaged distribution, with higher
positive peak amplitudes and smaller negative values. The last presented
time-step at around $t = 7$ fm/c is still strongly fluctuating
for the relativistic case.
The weighted vorticity averaged over all layers parallel to
the reaction, [x-z] plane at
different time steps are shown in
Figs. \ref{figG4-al}, \ref{figG84-al}, \ref{figG164-al}.
The dominant effect of the relativistic treatment is visible
the most in Fig. \ref{figG164-al}. Comparing eqs. \ref{wc} and \ref{wt}
the role of the relativistic $\gamma$ factor is apparent. As discussed
after eq. (\ref{wt}) the large peripheral velocities make the vorticity
large at the external surfaces of the expanding system. At these
external regions the matter density is small and the relative density
fluctuations are large, so the fluctuations of the surface region
vorticities are large.

\section{The properties of vorticity and circulation}

In section \ref{vorticity} we introduced the classical vorticity and
circulation with the conservation laws for the circulation in case
of certain conditions.
These can be extended to the relativistic case if we define the
relativistic circulation \cite{Florkowski2010} (chapter 14.3) as:
$$
\Gamma(C)=\frac{1}{m}\oint_C w\, u^{\mu}\delta x_{\mu},
$$
where the weight, $w$, is the specific
enthalpy, $e+P$, over the baryon number, $n$,
and $m$ is the effective mass per net baryon. Applying the Euler
equation for perfect fluids results in
\be
\frac{\partial \Gamma(C)}{\partial \tau} =
\frac{1}{m} \oint_C \frac{\partial^\nu P}{n} \delta x_{\nu} +
\frac{1}{m} \oint_C w\, u^{\mu}\delta u_{\mu} \ ,
\ee
where the last term vanishes along the flow stream line because
$u^{\mu}\delta u_{\mu} = 0$. Thus, if the pressure, $P$ depends on
the density, $n$, only the second term for a closed loop integral
also disappears, and the circulation remains constant
just as in the classical case.

In our problem these conditions are not satisfied, so we used the
same weighting as in the classical case, and evaluated the circulation
with the weighting as in the classical case to enable us the comparison,
see Fig. \ref{Gamma-cl}. We performed the integral over the [x-z]
surface for the weighted relativistic vorticities avaraged over all
$y$-layers.

Let us take the surface area of the reaction plane, which is filled with
fluid and take the bounding curve at the outside edges of the
fluid in this plane. This curve expands with the fluid.
In our case in rapidly expanding and non-perfect fluid the circulation
as well as the vorticity in in the [x-z] reaction plane are both decreasing.
\begin{figure}[ht] 
\begin{center}
      \includegraphics[width=7.6cm]{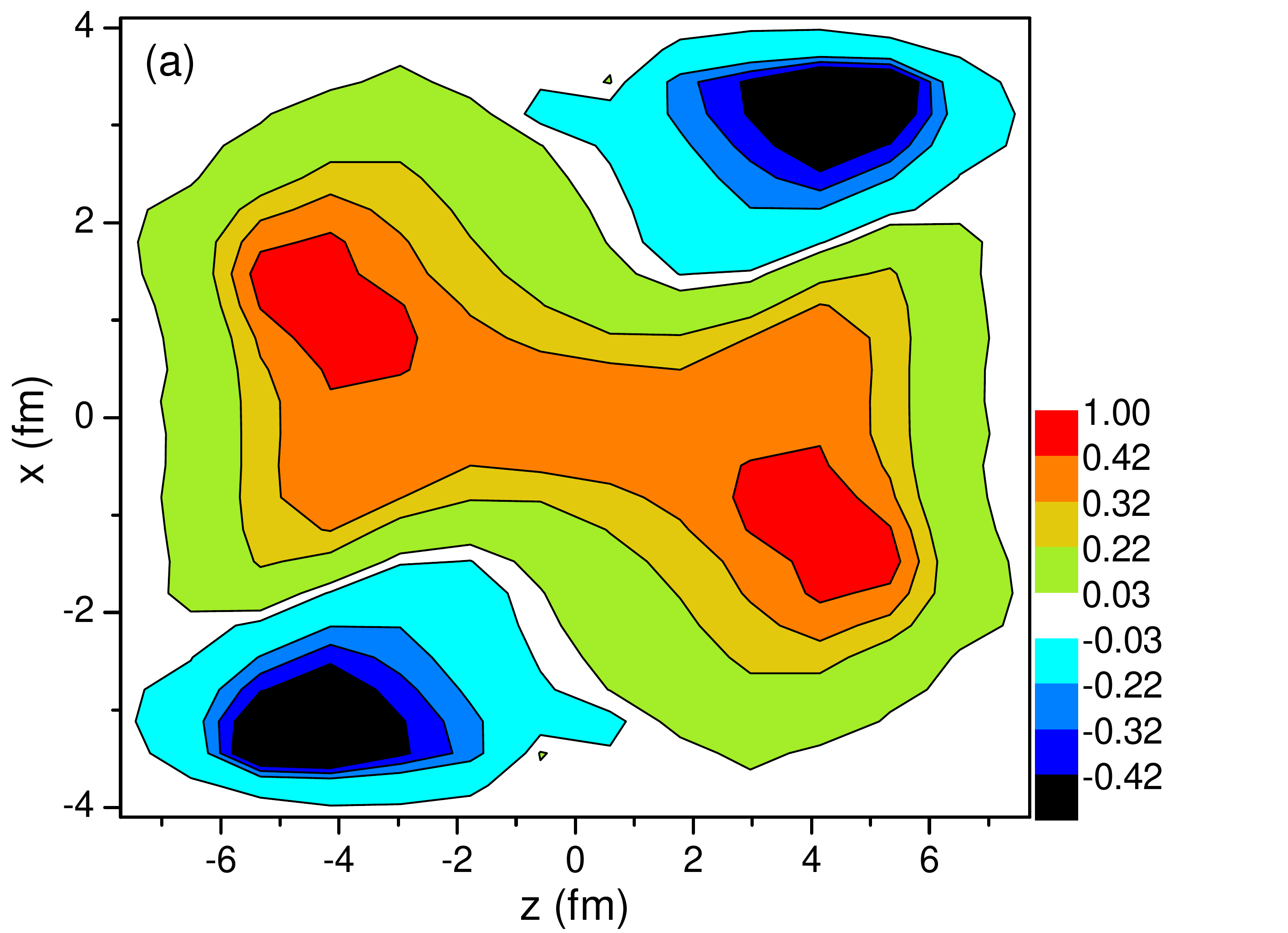}
      \includegraphics[width=7.6cm]{7b-G4-all-layer.pdf}
\end{center}
\vspace{-0.3cm}
\caption{
(Color online)
The weighted relativistic vorticity
calculated for all [x-z] layers at t=0.17 fm/c.
The collision energy is $\sqrt{s_{NN}}=2.76$ TeV,
$b=0.5\,b_{max}$ for (a) and $b=0.7\,b_{max}$ for (b).
The configuration (a) is not favoring KHI while (b) is.
The cell size is $dx=dy=dz=0.585/0.4375$ fm.  The average weighted
vorticity in the reaction plane is 0.07241/0.19004 c/fm for the two
cases respectively.
Notice the different color coding scales for (a) and (b).}
\label{figGHG4-al}
\end{figure}

For us the rate of this decrease is important, to see if we can
still detect the vorticity and circulation at freeze out.
Notice that we calculated only the $\Omega_{zx}$ component
of the weighted vorticity distributions.
Due to the close to spherical expansion, the direction of vorticity
may develop into different directions. This also contributes to
the decrease of the circulation.

\begin{figure}[ht] 
\begin{center}
      \includegraphics[width=7.6cm]{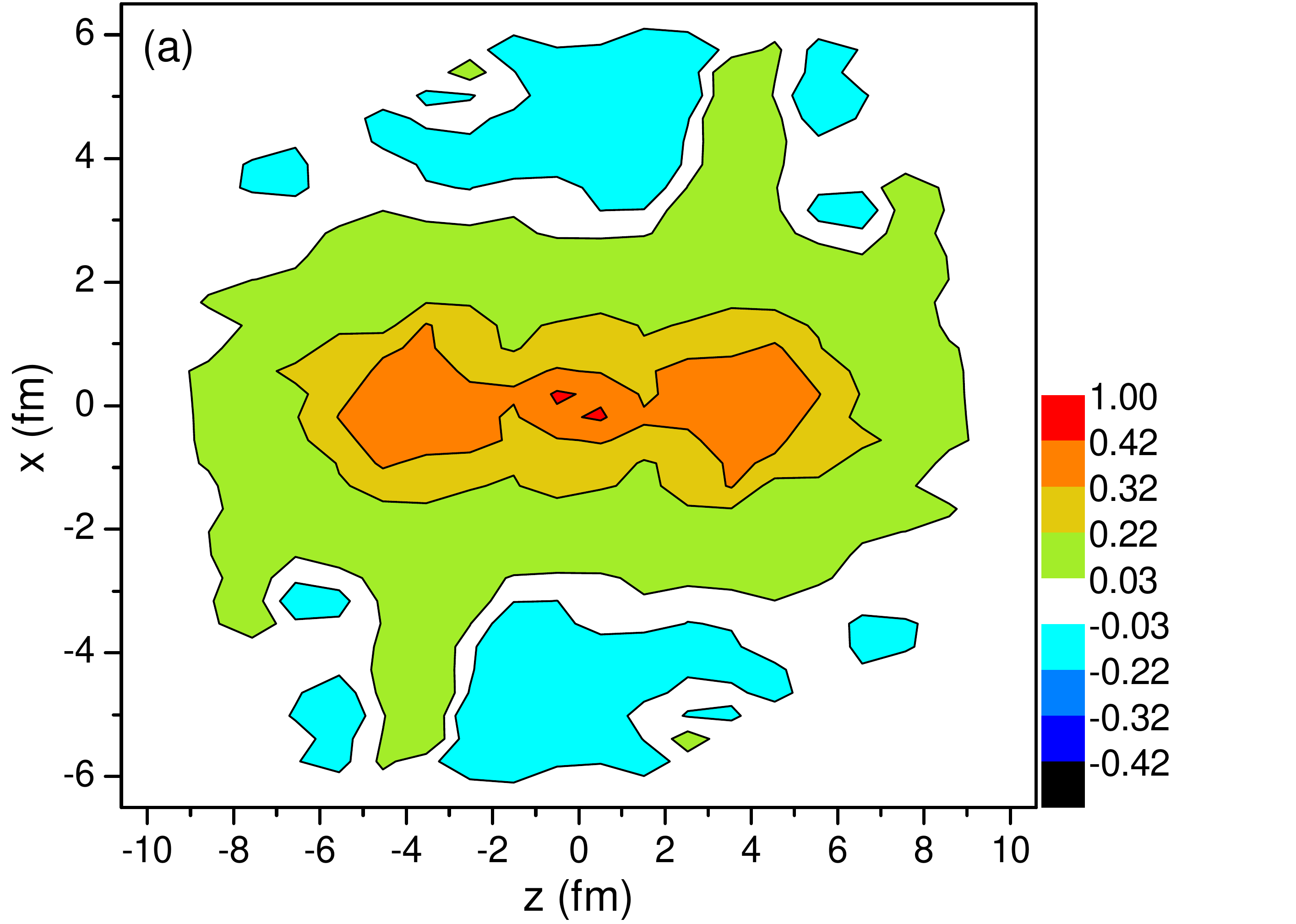}
      \includegraphics[width=7.6cm]{8b-G84-all-layer.pdf}
\end{center}
\vspace{-0.3cm}
\caption{
(Color online)
The weighted relativistic vorticity
calculated for all [x-z] layers at t=3.56 fm/c.
The collision energy is $\sqrt{s_{NN}}=2.76$ TeV,
$b=0.5\,b_{max}$ for (a) and $b=0.7\,b_{max}$ for (b).
The configuration (a) is not favoring KHI while (b) is.
The cell size is $dx=dy=dz= 0.585/0.4375$ fm.  The average vorticity in the
reaction plane is 0.05242 / 0.10685 c/fm for the two
weighted vorticity respectively.
Notice the different color coding scales.}
\label{figGHG84-al}
\end{figure}

We calculated and presented the weighted vorticity distribution
in the reaction plane, and then also calculated it for all the
[x-z] layers at different y-values, and took the average of these
vorticity distributions, see Figs.  \ref{figG4-al}, \ref{figG84-al},
\ref{figG164-al}, \ref{figGHG4-al}b, \ref{figGHG84-al}b.
The overall vorticity is positive, this originates from the  initial
shear flow configuration. It decreases with time, not only because the
weight of the reaction plane decreases due to the expansion, but also
because of the viscous dissipation in the 3D expansion.

The examples above are for the configuration where the
Kelvin-Helmholtz Instability (KHI) is predicted to occur in
heavy ion collisions at LHC.

We also studied, how the vorticity and circulation changes
with increased (numerical) viscosity and for more central
collisions, where the occurrence of KHI is not predicted
by the CFD model, see Figs. \ref{figGHG4-al}a, \ref{figGHG84-al}a.

In ref. \cite{hydro2} it was suggested that the KHI leads
to an increased $v_1(y)$ peak. We search for other,
preferably more sensitive and more specific experimental methods
to identify Rotation and KHI, and possibly
also separate the two effects.

The vorticity is also strikingly different for the configurations
which are adequate for KHI, and for those which not, and show
only rotation.
The initial and intermediate time stages are compared for
two different configurations where KHI may and may not appear.

The primary reason of the difference with KHI is that in the highly
peripheral reactions the profile height is smaller but the
asymmetry between the target and projectile contributions
is larger at the edges, and thus the shear in the matter is
considerably larger. Consequently both the maximum value
of vorticity and the average are larger for the more peripheral
configuration: the initial average vorticity is almost 3-times
larger in the more peripheral configuration favourable for KHI
than in the less favorable one.

At the later time, $t=3.56$ fm/c, the difference is still large,
the KHI formation leads to a vorticity, which is about a factor
two larger,  Fig. \ref{figGHG84-al}.

This, strong dependence on impact parameter arising in
low viscosity matter is very promising from the point of view of the
observability of the effect.

The possibility for observation mentioned in ref. \cite{hydro2},
by the position of the rotated $v_1(y)$ peak, is also visible
in Fig. \ref{figGHG84-al} where the vorticity peak position is seen
in the two configurations. This peak position may be coupled to the
peak position of the earlier mentioned $v_1$-peak. Here in the
configuration favorable for KHI the forward rotation angle of the
peak is $37^o$, while in the less favorable configuration it is $33^o$.
Thus, the KHI arising only in low viscosity matter leads to a
special increase in the rotation and vorticity. The observability
via finding the motion of the collective (y-odd) $v_1(y)$ peak is not
easy in present LHC experiemnts because the initial state fluctuations
contribute to large longitudinal c.m. fluctuations, apart of azimuthal
fluctuations in the transverse plane \cite{hydro1}.

\section{Conclusions}

An analysis of the vorticity and circulation development was performed
for peripheral
Pb+Pb reactions at the CERN LHC energy of $\sqrt{s_{NN}}=2.76$ TeV.
The initial peak vorticity was more than 10 times larger (exceeding 3 c/fm)
than the one obtained from random fluctuations in the transverse
plane, of about 0.2 c/fm \cite{Stefan}.
The reason is in the high initial angular
momentum arising from the beam energy in non-central collisions.

Although the vorticity and circulation decreases rapidly due to the
explosive expansion of the system still at 4 fm/c after the
beginning of fluid dynamical expansion the peak vorticity is above
3 c/fm in favorable configurations with KHI development and
it reaches up to 1 c/fm at the same time for less favorable
configurations.

This makes it promising to observe the consequences of this
rotation and its sensitivity to turbulent configurations.
Not only the peak vorticity exceeds earlier estimates
from random fluctuations,
but also the predicted average vorticity is substantial, it reaches
0.2 c/fm at favourable initial configurations with KHI, while the
average vorticity originating from random fluctuations is vanishing
\cite{Stefan,Oleg}.
The estimated angular deflection arising from this random origin
"chiral vortaic effect (CVE)"  \cite{Oleg},
is small, $\cos(\delta \phi) \sim 10^{-5}$. The effect arising from
the initial angular momentum of a peripheral collision is bigger.
This improves the observability of the average vorticity compared
to CME and CVE predictions.

In this work we just repeat the earlier mentioned observable signatures
related to the (rapidity odd component) of the directed, $v_1$ flow,
which is a promising possibility for the observations \cite{Eyyubova}.
The time sequences of the results show that the maximum of vorticity
in the side regions rotates, with time it moves forwards on the top,
the maximum reaching the positive $z$-side for the latest times presented.
The even flow harmonics are much less sensitive to this change. The
rapidity width of the $v_2(y)$ may be effected weakly.
Other methods in different correlation observables are more directly
connected to rotation in the flow and
will be addressed in forthcoming publications, (e.g. \cite{CV13}).

The question arises, how the surface energy influences the rotation
and the KHI. The external surface of the expanding QGP is significant,
as in the interior the quark gluon fluid has weak interaction
(asymptotic freedom) and small viscosity. However, the surface energy
has the strongest effect on the hadronization of QGP as first
described in ref.
\cite{CK92}.
On the other hand the collective flow, as well as the KHI, develops
primarily in the early QGP phase as the approximate quark number
scaling indicates. At RHIC and LHC energies, at this early stage
the surface energy is negligible compared to the energy of the
collective flow, and it primarily hinders the early emission of
low energy hadrons from the plasma, while it does not hinder the
rotation. In the KHI the situation is more involved. In the cases
where the KHI develops between two fluids (e.g. water/air or
oil/air) the large surface tension (e.g. between oil and air)
can hinder the development of KHI, as well known to sailors for
centuries. If the KHI develops within one fluid (like in air or
QGP), due to the large shear between the two fluid layers, there is no
surface tension in the conventional sense, but the layer with the
high shear may have extra energy, and so it can lead to
an effective surface tension,
which can weaken the development of KHI also. See estimates in ref.
\cite{WNC13}.

\begin{acknowledgments}

Enlightening discussions with Elena Bratkovskaya and Francesco Becattini are gratefully
acknowledged.

\end{acknowledgments}


\end{document}